\documentclass[10pt]{article} 
\usepackage{amssymb}
\usepackage{graphicx,epsf,rotate,subfigure} 
\usepackage{times,cite,color,xspace}
\usepackage{multirow}
\usepackage{amsmath}
\usepackage{tikz}
\newcommand{\met}{\ensuremath{\not\!\!E_T}\xspace}

\newcommand{\tr}[1]{{\color{black} #1}}

\begin{document}
\begin{flushright}
IFT-UAM/CSIC-14-037
\end{flushright}
\begin{center}
{\Large \bf The Higgsino-Singlino World at the Large Hadron Collider} \\
\vspace*{1cm} \renewcommand{\thefootnote}{\fnsymbol{footnote}} { {\sf
   Jong Soo Kim${}^{1}$}\footnote{email:jong.kim@csic.es}
   and {\sf Tirtha Sankar Ray${}^{2}$}\footnote{email:
  tirthasankar.ray@gmail.com}}
\\
\vspace{10pt} {\small ${}^{1)}$ {\em Instituto de Fisica Teorica
    UAM/CSIC, C/ Nicolas Cabrera 13-15, \\ Universidad Autonoma de
    Madrid, Cantoblanco, Madrid 28049, Spain} \\ ${}^{2)}$ {\em ARC
    Centre of Excellence for Particle Physics at the Terascale, School
    of Physics, \\University of Melbourne, Victoria 3010, Australia}}
\normalsize
\end{center}

\begin{abstract}
We consider light higgsinos and singlinos in the next-to-minimal supersymmetric Standard Model at the
Large Hadron Collider. We assume that the singlino is the lightest supersymmetric particle and that 
the higgsino is the next-to-lightest supersymmetric particle with the remaining supersymmetric particles 
in the multi-TeV range. This scenario, which is motivated by the flavor and CP issues, provides a phenomenologically viable dark
matter candidate and improved electroweak fit consistent with the measured Higgs mass.
Here, the higgsinos decay into on (off)-shell gauge boson and the singlino. We consider
the leptonic decay modes and the resulting signature is three isolated leptons and missing transverse energy which 
is known as the trilepton signal. We simulate the signal and the Standard Model backgrounds and present the 
exclusion region in the higgsino-singlino mass plane at the Large Hadron Collider at $\sqrt{s}=14$ TeV for
an integrated luminosity of 300 fb$^{-1}$.
\end{abstract}

\setcounter{footnote}{0}
\renewcommand{\thefootnote}{\arabic{footnote}}


\section{Introduction}
Supersymmetric (SUSY) models are very popular among the numerous TeV
extensions of the Standard Model (SM)
\cite{Haber:1984rc,Martin:1997ns}. One of the main tasks of the LHC is
the direct search for SUSY particles. After three years of running, both
LHC experiments ATLAS and CMS have not revealed any new particles
beyond the SM, but the absence of any excesses above the SM
expectation can be translated into strict limits on the parameter
space of low-energy SUSY. In particular, the first two generation
squarks and gluinos with masses below 1.7 TeV are excluded if the
squarks and gluinos are mass degenerate \cite{Aad:2014wea}.

This result together with a relatively heavy Higgs
\cite{:2012gk,:2012gu} somewhat undermines the rationale for TeV scale SUSY,
since heavy SUSY particles seem to reintroduce
finetuning. However, electroweak finetuning arises from the
minimization of the scalar potential.
The matching condition for electroweak symmetry
breaking is $1/2\,M_Z^2 \approx -\mu^2 -M_{H_u}^2$  \cite{Drees:1995hj}\footnote{Here, we
 assume $\tan\beta\ge5$.}, where $M_Z$ is the mass of the Z boson and
$M_{H_u}$ is the mass of the Higgs boson that couples to the top.
A very large $\mu$ term is unnatural due
to the required precise cancellation between the soft breaking terms
and $\mu$. Thus, a supersymmetric model with heavy multi-TeV scalars,
but sub-TeV $\mu$ values can still avoid large electroweak finetuning.

SUSY models with heavy multi-TeV matter scalars have the advantage
that loop induced flavor changing neutral current and CP violating
processes are suppressed
\cite{Gabbiani:1996hi,Dine:1990jd,Altmannshofer:2013lfa} and help to
ameliorate the late time gravitino decay problem \cite{Khlopov:1984pf,
  Kawasaki:2008qe}. Moreover,
potential baryon number violating dimension five operators are
suppressed and the resulting proton decay rate becomes very small
\cite{Murayama:2001ur}. Ref. \cite{Baer:2011ec} considers such a split
scenario with light and degenerate higgsinos and decoupled gauginos
and matter scalars ({\it higgsino world
  scenario}). However, in split scenarios with a light higgsino LSP the
annihilation cross section is too large. Assuming standard cosmology,
the thermal relic density is too small compared to the WMAP and Planck
measurement \cite{Hinshaw:2012aka,Ade:2013zuv}
$\Omega h^2\approx0.1187.$

The simplest extension of the MSSM is the next-to-MSSM (NMSSM) with a
scale invariant superpotential \cite{Ellwanger:2009dp}. The
supersymmetric Higgs mass term $\mu$ is dynamically generated by the
vacuum expectation value (vev) of a gauge singlet chiral superfield
$S$ and thus the NMSSM provides a weak scale solution of the $\mu$
problem in the MSSM. The singlet superfield leads to additional
singlet-like CP-even and CP-odd Higgs states as well as a singlino-like 
neutralino state and thus the additional degrees of freedom can
provide a solution to the dark matter issue of the {\it higgsino world
  scenario}. However, the resulting relic density is either too large
or to small in large region of parameter space in the singlet extended
{\it higgsino world scenario}. One solution is to demand a singlino-like 
neutralino LSP whose annihilation cross section is resonantly
increased via Higgs bosons in the $s$-channel. Another solution is
co-annihilation with a slightly heavier higgsino-like next-to-LSP
(NLSP). Both mechanism lead to the desired relic density and hence the
singlet extension of the {\it higgsino world scenario} is a
phenomenologically viable model \cite{Hugonie:2007vd,Belanger:2005kh}.

In this paper, we consider a {\it higgsino-singlino world scenario}
with multi-TeV matter scalars and decoupled gauginos, but with a small
$\mu$ term and a singlino-like LSP  i.e.
 $m_{\rm singlino}< m_{\rm higgsino} \ll m_{\rm scalar},\,m_{\rm gaugino}$.
 We want to explore the discovery reach of our scenario at the LHC
 at $\sqrt{s}=14$ TeV in the production of a neutralino-chargino pair,
\[ pp\rightarrow\tilde\chi_1^\pm\tilde\chi_{2,3}^0~~\mbox{with}~~
\tilde\chi_1^\pm\rightarrow W^{\pm(*)}\tilde\chi_1^0\quad\rm{and}\quad \tilde\chi_{2,3}^0\rightarrow Z^{0(*)}\tilde\chi_1^0.\]

The hadronic decays of the higgsinos lead to a final state signature
with large QCD background and thus is not a viable signal at the
LHC. However, the leptonic decay mode has particularly small QCD and
SM background. The signature is three isolated leptons and missing
transverse energy. This process is known as the trilepton signal and
the corresponding searches has been performed by ATLAS and CMS
\cite{ATLAS:trilepton,Aad:2014nua,Khachatryan:2014qwa}. Studies of the
discovery potential at 14 TeV has been studied in
\cite{Baer:1994nr,Baer:1995va} and in Ref.~\cite{Bornhauser:2011ab}
the discovery potential of CP violation in the trilepton channel has
been investigated. In this paper, we want to re-analyze the trilepton
study. We simulate the signal and background at hadron level and we
take into account the most important detector effect by performing a
fast detector simulation. In particular, we derive limits for higgsino-like charginos and 
neutralinos with a singlino-like LSP at the LHC at 14 TeV which has not been considered in
previous works.

The remainder of the paper is organized as follows. In
Sect.~\ref{sec:spectrum}, we discuss our scenario in more detail. In
Sect.~\ref{sec:gauge}, we briefly review the main phenomenological
features of the scenario. In Sect.~\ref{sec:LHC}, we
first discuss the constraints from LEP2 and the LHC8 results and then
the selection cuts before showing the numerical results for two
benchmark points. Finally, we show the discovery reach in the
higgsino-singlino mass plane at the LHC at $\sqrt{s}=14$ TeV for an
integrated luminosity of 300 fb$^{-1}$. We conclude in
Sect.~\ref{sec:conclude}.

\section{The Spectrum}\label{sec:spectrum}
We consider the scale invariant NMSSM \cite{Ellwanger:2009dp}. 
Assuming that the
gauginos and the sfermions with masses in the multi-TeV scale
are essentially decoupled from the low energy scale theory, we are left with the
following particle spectrum beyond the SM fields : (i) neutralinos: a
singlino-like LSP $(\tilde\chi^0_{1}),$ two higgsino-like neutralinos
$(\tilde\chi_2^0,\tilde\chi_3^0);$ (ii) charginos: higgsino-like state
$\tilde\chi^{\pm}_1;$ (iii) CP-even Higgs fields: the singlet like
field $(H_1),$ the SM like Higgs field $(h)$ and the heavy doublet-like 
CP-even scalar field $(H_2)$ and (iv) the CP-odd scalars: a
singlet-like scalar $(a)$ and a heavy CP-odd scalar $(A).$ In this
limit the entire effective theory of the \textit{higgsino-singlino
world scenario} essentially reduces to the following superpotential
and the corresponding soft breaking part of the Lagrangian assuming a $Z_3$ symmetry 
\begin{eqnarray}
 W_{(\mathcal{H\,S})} & = & \lambda \mathcal{S} \mathcal{H}_u\cdot \mathcal{H}_d +
 \frac{1}{3}\kappa\,\mathcal{S}^3, \nonumber\\ -\mathcal{L}_{\rm
   soft}^{(HS)} & = & m_{H_u}^2|H_u|^2 + m_{H_d}^2|H_d|^2 + m_S^2|S|^2
 + \left( \lambda A_{\lambda} H_u\cdot H_d S + \frac{1}{3} \kappa
 A_{\kappa}S^3 + {\rm h.c.} \right),
\label{spnmssm}
\end{eqnarray}
where $\mathcal{S}$, $\mathcal{H}_u$ and $\mathcal{H}_d$ denote the
singlet, SU(2) doublet up-type and the doublet down-type Higgs
superfields, respectively.  $S$, $H_u$ and $H_d$ are the respective
scalar fields. $\lambda$ and $\kappa$ are dimensionless Yukawa
couplings, whereas the soft breaking terms for the scalar fields are
given by $m_{H_u}^2$, $m_{H_d}^2$ and $m_S^2$. $A_\lambda$ and
$A_\kappa$ are the trilinear soft breaking terms. Once the singlet gets a vev,
the Higgs mixing term $ \mu \equiv \lambda \langle s\rangle $ is
generated.  This can easily be at the weak scale, solving the usual
$\mu$ problem of the MSSM.  In the remainder of this section, we briefly sketch the 
masses of the relevant sub-TeV particles in the theory.

In the
limit where $|\mu| \ll M_{\rm{gauginos}},$ the neutralino mixing
matrix block diagonalizes into the predominantly heavy gaugino sector
and the light higgsino-singlino sector. The mass matrix of the light higgsino-singlino sector can
be written as,
\begin{equation}
  {\mathcal{M}}_{\tilde\chi^0}=
\begin{bmatrix}
               0&-\mu & -\lambda v_u  \\
               -\mu & 0 & -\lambda v_d \\
                -\lambda v_u & -\lambda v_d & 2\kappa \langle s\rangle
\end{bmatrix}.
\label{neutralino}
\end{equation}
In this paper we only consider the parameter region with $2 \kappa < \lambda \ll 1.$  This choice ensures that the
lightest neutralino is predominantly singlino-like and
the chargino and neutralino masses are approximately
given by,
\begin{eqnarray}
 M_{\tilde\chi_{2,3}^{0}}= M_{\tilde\chi^{\pm}} \sim \mu = \lambda
 \langle s\rangle, \nonumber \\ M_{\tilde\chi^{0}_1} \sim 2 \kappa
 \langle s\rangle = 2\frac{\kappa}{\lambda} M_{\tilde\chi_{2,3}^0}.
 \label{neutralinomass}
 \end{eqnarray}

The Higgs sector is composed of the usual CP-even scalar state that
will be identified with the Higgs state observed at the LHC with a
mass around $\sim 125$ GeV. The mass of this state can be written as,
\begin{equation}
 m_h^2 = M_Z^2\left(\cos^22\beta + \frac{\lambda^2}{g^2}\sin^22\beta
 \right) + \delta,
\end{equation}
with $\delta$ quantifying the radiative contributions from the
sparticles, mainly from the stops. It will be assumed that the masses
of the heavier sfermions will be set by fixing the Higgs mass at 125
GeV.\footnote{The sfermion mass spectrum depend on the mixing in the
  stop sector and can be quite light. However in this paper we will
  focus on models where the mixing is negligible leading to a
  decoupled stop sector. Realistic origins of such spectrum will be
  given elsewhere in the discussion.} The masses of the other light
singlet-like CP-even and CP-odd Higgs is given by,
\begin{eqnarray}
 m_{H_1}^2 & \sim & \kappa \frac{M_{\tilde\chi_{2,3}^0}}{\lambda}
 \left(A_{\kappa} + 4 \kappa \frac{M_{\tilde\chi_{2/3}^0}}{\lambda} \right),
 \nonumber \\ m_{a}^2 & \sim & -3 \kappa
 \frac{M_{\tilde\chi_{1/2}^0}}{\lambda} A_{\kappa}.
\end{eqnarray}
The sub-TeV spectrum will also include the usual doublet type CP-even
($M_{H_2}$) and CP-odd Higgs ($M_A$) which have nearly degenerate mass
given by,
\begin{equation}
 M_{A} =
2 \mu (A_{\lambda} +\kappa \langle s\rangle)/\sin2\beta,
\end{equation}
and the charged Higgs with mass $M_{H_{\pm}}^2 = M_A^2 +M_W^2.$ In the following, we choose $\lambda \ll 1$,
  so that the singlino and the singlet-like Higgs bosons only couple weakly to the other
  particles and thus possible light singlet-like Higgs bosons are not excluded by the LEP constraints. As we discuss below this is also the 
  parameter range that is consistent with the Dark Matter constraints. 

This framework is conceptually different from the split \cite{Giudice:2004tc, ArkaniHamed:2004fb, Demidov:2006zz} or mini-split \cite{Arvanitaki:2012ps}
models due to the existence of additional light scalar states. This
necessarily includes additional sources of fine-tuning in the classical sense. However, in the paradigm where this notion of naturalness
is disregarded \cite{Dubovsky:2013ira, Susskind:2003kw} or reformulated \cite{Wells:2004di, Farina:2013mla, deGouvea:2014xba}, such proliferation of sources of fine-tuning may not be considered as conceptually inconsistent. Indeed, in models
where the Higgs-Singlet sector is sequestered from the rest of the supermultiplets, 
this kind of spectrum can naturally arise. E.g., in 5d SUSY models where the Higgs-Singlet multiplets are usually 
confined to the brane and the rest of the
multiplet can access the bulk \cite{Bhattacharyya:2012ct}. 

\section{Phenomenology of the Higgsino-Singlino World Scenario}\label{sec:gauge}
In this section we briefly comment on some phenomenological aspects of the {\it higgsino-singlino world scenario}:
\begin{figure}
\begin{minipage}[t]{0.47\textwidth} \centering
\begin{tikzpicture}
\node (plot) {\includegraphics[height=.8\textwidth, width=.8\textwidth,angle=0]
             {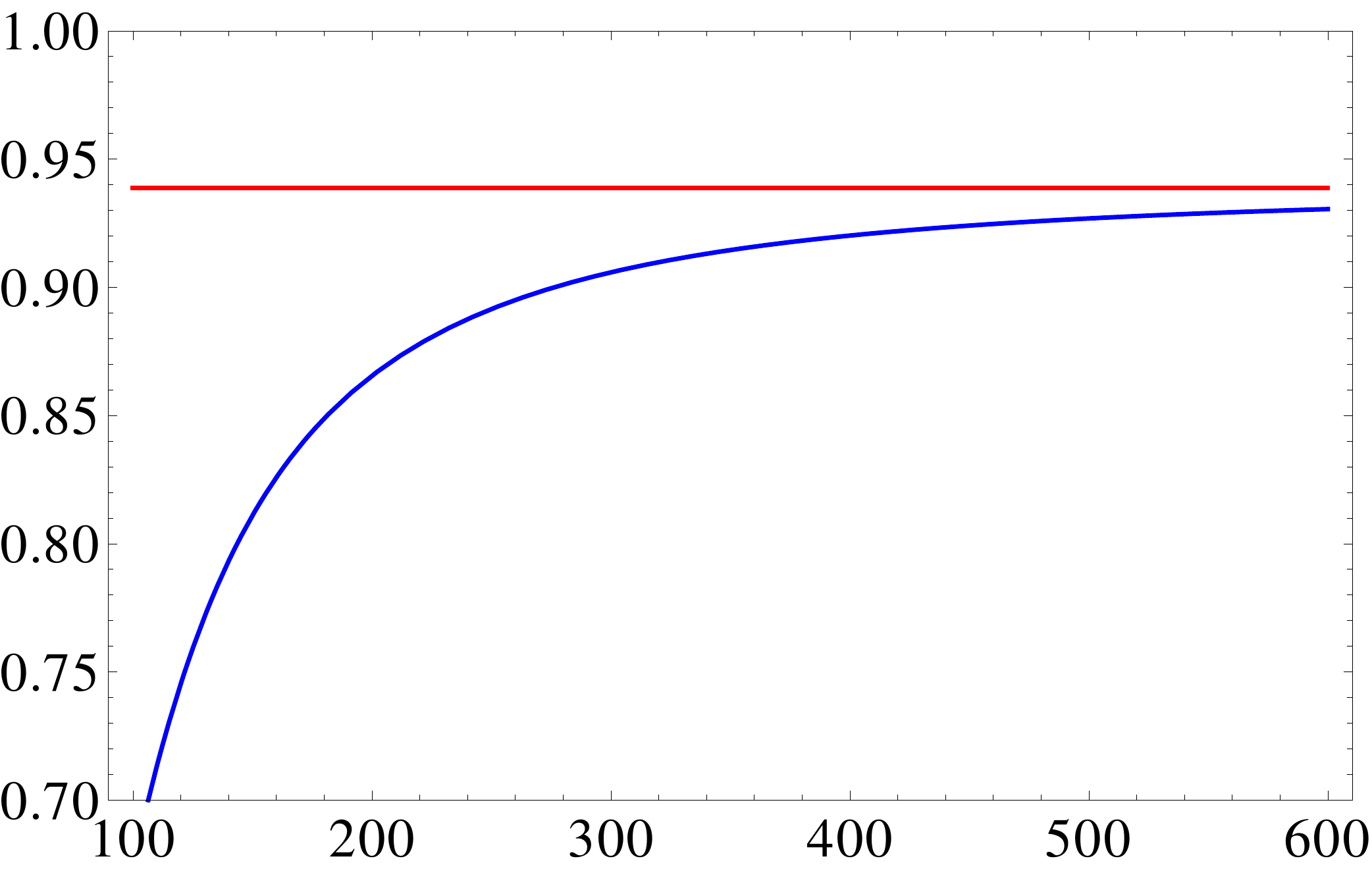}};
\node [rotate=90] at (-4,.5) {\Large{$\mathbf{\chi^2/d.o.f.}$}};
\node at (0,-4) {\Large{$\mathbf{M_{\tilde{\chi}^0_2}~ [GeV]}$}};
\end{tikzpicture}
	   \caption{\em \small  $\chi^2/d.o.f. $ fit for the
                 \textit{higgsino-singlino} world scenario using the three observable $M_W, s^2_{l}$
                 and $\Gamma(Z\rightarrow l^+l^-)$ is shown in blue.  The horizontal red
                 line represents the SM prediction.}  
             \label{electroweak}
\end{minipage}
\hspace{5mm}
\begin{minipage}[t]{0.47\textwidth} \centering
\begin{tikzpicture}
\node (plot) {\includegraphics[height=.8\textwidth,
    width=.8\textwidth,angle=0] {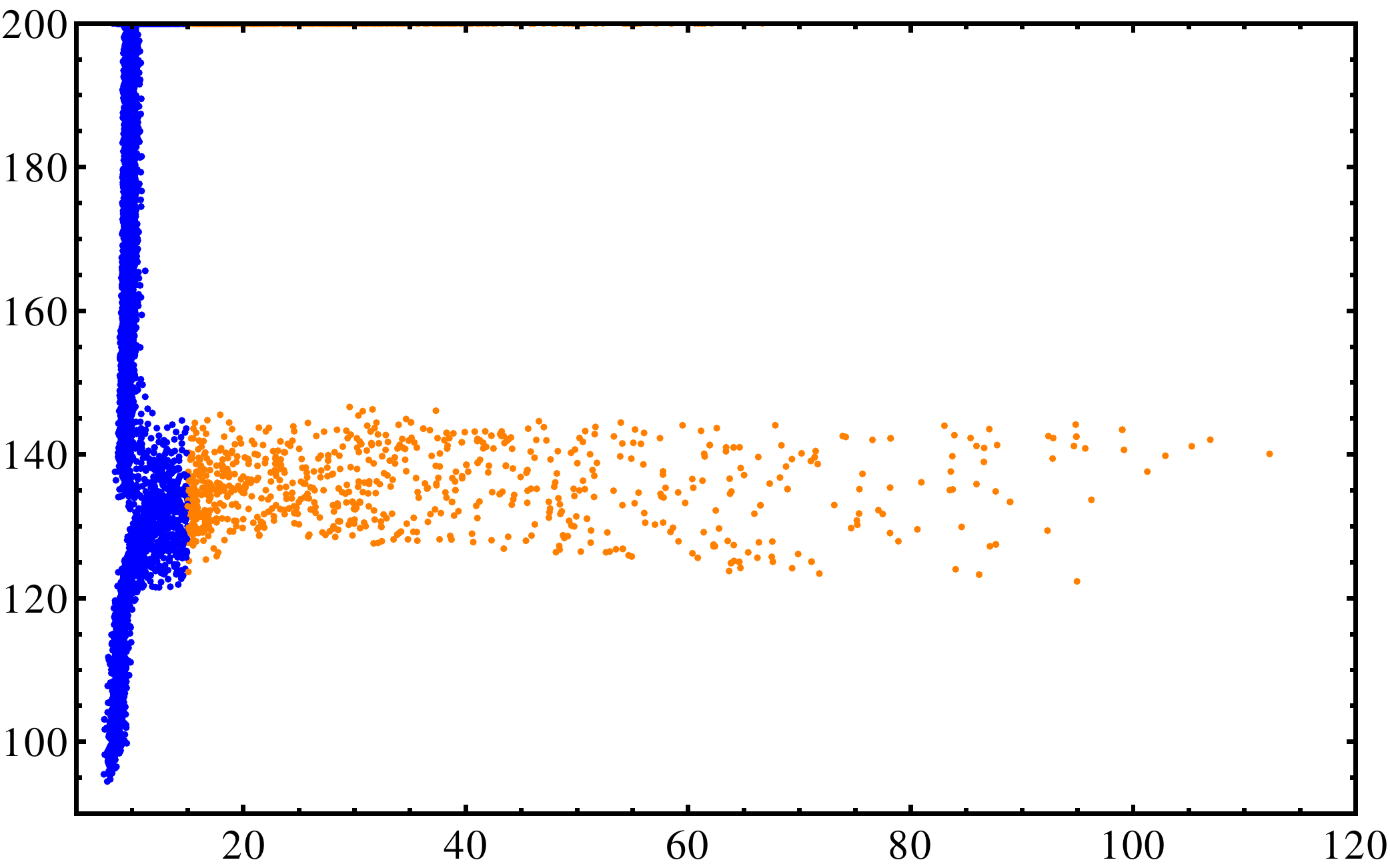}}; \node [rotate=90] at
(-4,.5) {\Large{$\mathbf{M_{\tilde{\chi}^0_1}~ [GeV]}$}}; \node at
(0,-4) {\Large{$\mathbf{\left(
      M_{\tilde{\chi}^0_2}-M_{\tilde{\chi}^0_1}\right)~ [GeV]}$}};
\end{tikzpicture}
\caption{\em \small Scatter plot of allowed
                       parameter points with $M_A= 300$ GeV is
                       displayed.   The
                 blue (darker) points represent the 
                 co-annihilation region while the red (lighter)
                 points represent regions in the parameter space
                 where the relic abundance is obtained through
                 resonant s channel annihilation of the singlino
                 LSP.}
                 \label{DarkMatter}
\end{minipage}
\end{figure}
\begin{enumerate}
 
\item With this split spectrum one can achieve a  slightly better fit to the electroweak precision observables. 
Assuming $\lambda \ll 1$ one can neglect the effect of the
singlet-doublet mixing. With this assumption there is negligible contribution
to the S and T parameters\footnote{There are however
  small corrections that arise due to the mixing with the singlino
  \cite{Barbieri:2006bg}.An analytical expression for the
  contribution is difficult. A perturbative expansion of the mass
  matrix given in Equation~(\ref{neutralino}) in terms of $\lambda$
  gives us $\Delta T \sim \lambda^4 v^2/\mu^2 \cos 2\beta
  f(M_{\tilde\chi^{0}_1}/\mu).$ In the limit $\lambda \ll 1$
  contributions of this order can be safely ignored.}\cite{Martin:2004id}. The non-zero
contributions can be parametrized using the three observables $M_W,
s^2_{l}$ and $\Gamma(Z\rightarrow l^+l^-).$ The latest experimental
values are given in \cite{Beringer:1900zz}. We utilize the SM prediction
including all computed higher order corrections for $M_W$
\cite{Awramik:2003rn}, the leptonic weak mixing angle $(s^2_{l})$ \cite{Awramik:2006uz} and
$\Gamma(Z\rightarrow l^+l^-)$ \cite{Freitas:2014hra}. The fit to the
experimentally measured values is slightly improved in a large region
of the parameter space as shown in Figure~\ref{electroweak}.
\item 
In the limit $2\kappa < \lambda\ll1$, that we explore in this paper, the singlino-like 
neutralino is the lightest supersymmetric particle, see
Equation~(\ref{neutralinomass}).  With conserved R-parity this can be
the dark matter candidate in this class of models \cite{Belanger:2005kh,Barger:2005hb}. 
The higgsino-like
neutralinos and charginos are the NLSPs. They are essentially
degenerate with mass $\sim\mu$, except the electroweak corrections
that lifts the degeneracy making the charginos slightly heavier by
$\sim {\mathcal{O}}(10)$ MeV.  

We have performed a systematic scan to obtain the region of parameter space
that is consistent with the dark matter relic density. Considering that the sfermions and the gauginos are decoupled,
one finds
that the entire parameter space of the theory, as expressed in Eq.~\ref{spnmssm},  can be defined in terms of
the following parameters, $ [\lambda,\ \kappa,\ \tan\beta,\ 
\mu_\mathrm{eff},\ A_\kappa,\ M_A].$ We utilize {\tt NMSSMTools}
\cite{Ellwanger:2004xm,Ellwanger:2005dv}and {\tt micrOmegas}
\cite{Belanger:2013oya} to perform a scan over in the  range:
\[ \lambda = [0.001, 0.1] ,\ \kappa=[-0.05,-0.01],\ \tan\beta =[1.5,20],\ 
\mu_\mathrm{eff} =[100,300]~GeV, \ \ A_\kappa=[1,1000]~GeV. \]

A scatter plot for points consistent with observed relic density is shown in Figure~\ref{DarkMatter}  for $M_A = 300$ GeV.
The allowed parameter space can be
divided into two distinct regions. In the  region of the parameter
space where $\kappa \sim \lambda/2,$  the right dark matter relic
density for LSP is achieved  through co-annihilation with a relatively
degenerate higgsino-like neutralino.  The LSP
can have a significant higgsino component in this scenario.
In this case for effective
reduction of the number density a mass difference of $\Delta M =
M_{\tilde\chi^0_1} -M_{\tilde\chi^0_2} < 20$ GeV is required.  The
relative degeneracy of the chargino and the neutralino makes it
relatively difficult to probe at the LHC. The collider phenomenology
of this region of the parameter space closely resembles the
\textit{higgsino world scenario} and have better prospects of being
probed at future colliders like the ILC \cite{Baer:2011ec}. For the
possibility of probing this region at LHC with mono-jets + \met, \tr{see}
\cite{Han:2013usa,Baer:2014cua,Schwaller:2013baa}.
A phenomenologically more promising region is obtained when
$M_{\tilde\chi^0_1} = 1/2 M_A,$ where $M_A$ is the mass of the heavy
CP-odd Higgs. In this case the LSP can have efficient resonant
annihilation with the heavy Higgs scalars in the $s$-channel. Actually
we observe that the relative degeneracy of the CP-even and CP-odd
heavy doublet-like Higgs implies a double resonance through the
process $\tilde\chi^0_1 \tilde\chi^0_1
\rightarrow~\mbox{on-shell}~H_2/A \rightarrow b\bar{b}, aa^*$. In this case the LSP is predominantly
singlino-like. In this case a
relative separation between the higgsinos and the singlinos of the
order of 100 GeV is possible. In the rest of this paper we will concentrate on the collider signal
of this region of the parameter space. 

\end{enumerate}

\section{Discovery Potential of Higgsino-Singlino World Scenarios at the LHC at $\sqrt{s}=14$ TeV}\label{sec:LHC}
In this section, we want to discuss the phenomenological consequences
of our \textit{higgsino-singlino} world scenario at the LHC with a relatively simple collider study. The
higgsinos and singlinos are the only kinematically accessible
supersymmetric states at the hadron collider. Here, we consider
associated chargino-neutralino pair production,
\begin{equation}
pp\rightarrow \tilde \chi_1^\pm\tilde \chi_2^0,\, \tilde \chi_1^\pm\tilde \chi_3^0.
\end{equation}
The cross section for chargino-neutralino pair production is
determined by $\tan\beta$, $\lambda$, the higgsino mass parameter and
the singlino mass.  Motivated by the LEP constraints on light singlet like scalars and the Dark Matter constraints discussed
above we set $\tan\beta=10$ and $\lambda=0.01$ in our study. However the results presented here is relatively
insensitive to the specific choice of these parameters. In particular, a different value of lambda only modify
the branching ratio of the higgsino slightly as long as lambda is small.
In Table \ref{tab:crosssection}, we show the total
chargino-neutralino pair production cross section in picobarn at the
LHC for $\sqrt{s}=14$ TeV \cite{Beenakker:1999xh}.  We assume that the
gauginos and sfermions are decoupled and we set $M_{H_{\pm}} >
M_{\tilde{\chi}_1^\pm} -M_{\tilde{\chi}_1^0},$ thus the $\tilde
\chi_1^\pm$ decays into $W^{(*)}\tilde\chi_1^0$ with a branching ratio
of $100\%$, where the asterisk denotes off-shell $W$ bosons. The
heavier neutralino eigenstates $\tilde\chi_2^0$ and $\tilde\chi_3^0$
generally decay into $Z^{(*)}\tilde\chi_1^0$.  However,
$\tilde\chi_2^0$ and $\tilde\chi_3^0$ can also decay into the CP-even
and CP-odd Higgs bosons.  The explicit decay properties depend on the
details of the Higgs sector.  We set $M_A > M_{\tilde{\chi}_2^0}
-M_{\tilde{\chi}_1^0},$ thus kinematically disallowing the
$\tilde\chi_2^0$ and $\tilde\chi_3^0$ to decay into the heavy doublet
like Higgs.  The branching ratios of the $\tilde\chi_2^0$ and
$\tilde\chi_3^0$ into singlet-like Higgs states are negligible, since
we consider $\lambda\ll 1$. The branching ratio of the neutral
higgsino states into the SM-like Higgs $h$ with a mass of $125\pm3$
GeV cannot be neglected, if the decay is kinematically possible.
However, the branching ratio of $\tilde\chi_2^0$ and $\tilde\chi_3^0$
into $Z$ is still sizable.

\begin{table}
\begin{center}
\begin{tabular}{c|c|c|c|c|c|c|c|c}
 $\mu$ [GeV] & 140 & 200 & 260 & 320 & 380 & 440 & 500 & 560\\ \hline
  $\sigma$ [pb] & 3.315 & 0.921 & 0.350 & 0.159 & 0.075 & 0.044 &
  0.026 & 0.016 \\
\end{tabular}
\caption{The total $\tilde\chi_{1}^\pm\tilde \chi_{2,3}^{0}$
  production cross section in picobarn at NLO at the LHC for
  $\sqrt{s}=14$ TeV \cite{Beenakker:1999xh}.}
\label{tab:crosssection}
\end{center}
\end{table}

We focus on the leptonic decay modes of the gauge bosons which results
in the trilepton and missing transverse energy final state
configuration. The trilepton and missing transverse energy (\met)
signal at the LHC was first investigated in
\cite{Baer:1994nr,Baer:1995va}. The ATLAS
\cite{ATLAS:trilepton,Aad:2014nua} and CMS \cite{Khachatryan:2014qwa}
searches for trilepton and large missing transverse momentum at the
LHC at $\sqrt{s}=8$ TeV with an integrated luminosity of 20 fb$^{-1}$
put already strict constraints on gaugino pair production in a
simplified MSSM model. They consider wino-like lightest chargino
$\tilde \chi_1^\pm$, heavier wino-like neutralino $\tilde\chi_2^0$ and
a bino-like LSP $\tilde\chi_1^0$ with decoupled sfermions and
higgsinos. $\tilde\chi_1^\pm$ and $\tilde\chi_2^0$ masses up to 345
GeV are excluded. However, the mass limits on charginos and
neutralinos are much weaker for the \textit{higgsino-singlino} world
scenarios, since the production cross section for higgsino-like
charginos and neutralinos are much smaller than for the
winos. Ref.~\cite{Ellwanger:2013rsa} published a study with a light
\textit{higgsino-singlino} scenario. They derived constraints in the
$M_{\tilde\chi_1^\pm}-M_{\tilde\chi_1^0}$ mass plane from the ATLAS
trilepton and \met search \cite{ATLAS:trilepton}. They found that
chargino masses up to 250 GeV are excluded. For small mass differences
between the higgsino and the singlino, searches from LEP for $e^+e^-
\rightarrow \tilde\chi_2^0\tilde\chi_1^0$ are relevant
\cite{Abdallah:2003xe,Abbiendi:2003sc}. As demonstrated in
\cite{Ellwanger:2013rsa} the LEP bounds are stricter than the current
LHC bounds for $M_{\tilde\chi_1^\pm}\leq140$ GeV. In the following, we
derive the exclusion limits in the \textit{higgsino-singlino} mass
plane at the LHC for an integrated luminosity of 300 fb$^{-1}$ at
$\sqrt{s}=14$ TeV.

The mass spectrum, couplings and decay widths are obtained with {\tt
  NMSSMTools 4.1.0} \cite{Ellwanger:2004xm,Ellwanger:2005dv, Belanger:2013oya}. The
signal events are generated with {\tt Herwig 2.7.0}
\cite{Bahr:2008pv}. The signal cross sections are normalized with the
next-to-leading order (NLO) calculation from {\tt Prospino2.1}
\cite{Beenakker:1999xh}. The dominant SM backgrounds $WZ$, $ZZ$ and
$t\bar t$ are generated with {\tt Herwig2.7.0}. The NLO cross sections
for vector boson pair production and $t\bar t$ are taken from {\tt
  MCFM 6.7} \cite{Campbell:2011bn} and \cite{Bonciani:1998vc},
respectively.  We have generated $5\times 10^{5}$ leptonic $WZ$,
$5\times 10^5$ leptonic $ZZ$ events and $10^{6}$ leptonic $t\bar t$
events.  The detector effects are estimated with the fast detector
simulation {\tt Delphes 3.0.12} \cite{deFavereau:2013fsa}.  We
replaced the ATLAS detector card of {\tt Delphes 3.0.12} with the {\tt
  CheckMATE 1.1.4} card\cite{Drees:2013wra}.  The detector tuning of
{\tt CheckMATE 1.1.4} has been validated with several ATLAS studies
(in particular with \cite{ATLAS:trilepton}) and hence should be more
accurate. Our event samples are then analyzed with the program package
{\tt ROOT} \cite{Antcheva:2009zz}.

Jets are defined using the {\it anti}-$k_T$ algorithm
\cite{Cacciari:2011ma} with $\Delta
R=\sqrt{(\Delta\Phi)^2+(\Delta\eta)^2}=0.4$.  Here, $\Delta\Phi$ and
$\Delta\eta$ are the difference in azimuthal angle and rapidity,
respectively.  We demand that all jets have $p_T>20$ GeV and
$|\eta|<2.5$. The $b$-tagging efficiency is 85$\%$. ATLAS
distinguishes between different {\it kinds} of electrons which have
different reconstruction and identification efficiencies as a function
of $\eta$ and $p_T$.  We require "tight" electrons in our
study\cite{Drees:2013wra}. All electrons must have $p_T>7$ GeV and
$|\eta|<2.5$. The electrons must be isolated, {\it i.e.}, the scalar
sum of the transverse momenta of the tracks within $\Delta R=0.3$ of
the electron must be less than 16$\%$ of the electron $p_T$
\cite{ATLAS:trilepton}.  As for the electrons, ATLAS also have
different types of muons with different efficiencies.  We require
"combined+standalone" muons in the following\cite{Drees:2013wra}. We
also demand that all muons have to have $p_T>7$ GeV and
$|\eta|<2.7$. The isolation requirements for the muons are similar to
the electron case, but with a ratio of 12$\%$ \cite{ATLAS:trilepton}.

For the overlap removal we use the following procedure
\cite{ATLAS:trilepton}.  Any jet within $\Delta R\le0.2$ of an
electron will be removed.  This cut prevents double counting, since
electrons are usually reconstructed as jets as well.  Since we do not
want to consider electrons and muons from heavy flavor decays within
jets, all electrons and muons within $0.2\le\Delta R\le0.4$ of a jet
will be removed.

We have implemented the lepton triggers from
\cite{ATLAS:trilepton}. The single electron or single muon triggers
require at least one electron or one muon with $p_T\ge25$ GeV.  The
symmetric di-muon trigger demands at least two muons with each $p_T
\ge 14$ GeV, while the asymmetric trigger requires $p_T \ge 18$ GeV
and $p_T \ge10$ GeV.  For the symmetric di-electron trigger, at least
two signal electrons are required to have $p_T \ge 14$ GeV, while for
the asymmetric electron trigger, we demand $p_T \ge 25$ GeV and $p_T
\ge 10$ GeV.  Finally, the mixed electron-muon (muon-electron) trigger
requires one electron with $p_T > 14$ GeV (10 GeV) and one muon with
$p_T \ge 10$ GeV (18 GeV). In the following, we assume an overall
trigger efficiency of 100 $\%$.

All events in the signal regions must contain three isolated leptons
(electrons and muons). We demand at least one same flavour opposite
sign (SFOS) lepton pair with an invariant mass above $20$ GeV to
suppress low mass resonances. We have defined three signal regions
with one $Z$ depleted region and two $Z$ enriched regions.

For the $Z$ depleted signal region {\bf SRnoZ}, we demand that the
SFOS pair closest to the $Z$ mass satisfies $m_{\rm SFOS}\leq81.2$ GeV
or $m_{\rm SFOS}\geq101.2$ GeV. Events with jets with $p_T\ge20$ GeV
are vetoed. Finally, we require $E_T^{\rm miss}\ge30$ GeV. This signal
region is very similar to the trilepton study presented in
\cite{Aad:2009wy}.

Both $Z$ enriched regions are defined as follows. We require for the
invariant mass $m_{\rm SFOS}$ closest to the $Z$ mass: $81.2\,{\rm
  GeV}\leq m_{{\rm SFOS}}\leq 101.2\,\rm{GeV}$. We veto all events
with $b$-jets with $p_T\ge20$ GeV.  We demand large missing transverse
energy with $E_T^{\rm miss}>75$ GeV and $150$ GeV corresponding to
signal regions {\bf SRZ1} and {\bf SRZ2}, respectively.  The
transverse mass is given by
\begin{equation}
m_T=\sqrt{2\cdot E_T^{\rm miss}\cdot p_T^\ell\cdot(1-\cos\Delta\phi_{l,E_T^{\rm miss}})},
\end{equation}
where $\Delta\phi_{l,E_T^{\rm miss}}$ corresponds to the azimuthal
angle between the lepton and the missing transverse momentum
vector. The lepton in the $m_T$ calculation is the one which is not
the lepton of the SFOS pair. $p_T^\ell$ is the transverse momentum of
the lepton. We demand $m_T\ge110$ GeV in order to suppress the $WZ$
background. {\bf SRnoZ} considers scenarios with small mass splitting
between the singlino and the higgsino, which is generally smaller than
the $Z$ mass. {\bf SRZ1} and {\bf SRZ2} target scenarios with larger
mass differences between the singlino and the higgsino. The difference
between {\bf SRZ1} and {\bf SRZ2} is the missing transverse energy cut
which is larger for {\bf SRZ2} and thus it is more sensitive for heavy
higgsinos and large mass differences between the higgsino and the
singlino.

We present the cutflows for the SM backgrounds as well as for two
benchmark points for an integrated luminosity of 300 fb$^{-1}$ at the
LHC with $\sqrt{s}=14$ TeV in Table \ref{tab:cutflowSRnoZ} and
\ref{tab:cutflowSRZ1}.  The statistical significance is estimated with
\begin{equation}
\mathcal{S}_{\rm stat}=S/\sqrt{B}, 
\label{eq:sig_stat}
\end{equation}
where $S$ and $B$ correspond to the number of signal events and background events after each cut.
We also show the significance taking into account the systematical errors.
We assume an overall systematic uncertainty of 10$\%$ for all SM backgrounds.
Our estimate of the significance is then given by
\begin{equation}
\mathcal{S}_{\rm stat\,+\,sys}=\frac{S}{\sqrt{B+(0.1\cdot B)^2}}.
\label{eq:sig_sys}
\end{equation}

First, we choose a light chargino with $M_{\tilde\chi_1^\pm}=160$ GeV
and a singlino with $M_{\tilde\chi_1^0}=100$ GeV for benchmark point
{\bf BP1}. Here both $\tilde\chi_2^0$ and $\tilde\chi_3^0$ decay via
off-shell $Z$ bosons. The first cut already provides a statistical
significance of 8.8 due to the large production cross section. The
dominant backgrounds are $WZ$ and $t\bar t$.  After the SFOS cut, we
veto $Z$ bosons and thus heavily suppress the $WZ$ and $ZZ$
background.  We apply a mild \met cut which reduces SM backgrounds
containing a $Z$.  However, nearly 20$\%$ of the signal events do not
pass the cut due to the small mass splitting between the higgsino and
the singlino. We keep this cut, since it removes the $Zb$ background
which we did not simulate \cite{Aad:2009wy}. The jet veto heavily
suppresses the $t\bar t$ background and we obtain a good statistical
significance of 13.6. Finally, if we account for systematic errors,
the significance reduces to 1.6 owing to the large systematic
uncertainty of the $WZ$ and $t\bar{t}$ backgrounds.  $t\bar t$ remains
as one of the dominant background in {\bf
  SRnoZ}. Note that our $t\bar t$ background is
larger than in \cite{Aad:2009wy}, partly because they did not normalise their
 $t\bar t$ sample to the NLO cross section. We rescaled their cross section 
 to NLO, but their $t\bar t$ background is still smaller than our estimate, because they 
 further reduced the $t\bar t$ background by imposing different isolation 
 criteria for the leptons. In addition, they demand a larger 
minimal transverse momentum of 10 GeV on the leptons. 
However, we keep our isolation
requirements, since we validated our analysis with
\cite{ATLAS:trilepton}. 
In any case, we believe that our background estimate for $t \bar t$ is 
sufficiently conservative.

In scenario {\bf BP2}, the neutralino and chargino masses are set to
$M_{\tilde\chi_1^\pm}=400$ GeV and $M_{\tilde\chi_1^0}=20$
GeV. On-shell decays of $\tilde\chi_{2,3}^0$ into $Z$ are still
dominant even though decays into the SM Higgs are kinematically
allowed.  The branching ratios are BR$(\tilde\chi_2^0\rightarrow
\tilde\chi_1^0+Z)=65\%$ and BR$(\tilde\chi_3^0\rightarrow
\tilde\chi_1^0+Z)=43\%$. The first two cuts are identical as for {\bf
  BP1}.  The SFOS and the $Z$ requirement suppress the $t\bar t$
background. We apply a $b$-jet veto which further reduces the $t\bar
t$ background while the other backgrounds are still sizable. However,
the strict cut on the missing transverse energy heavily suppresses the
di-gauge boson backgrounds. The final cut on $m_T$ further reduces
the $WZ$ background and we obtain a statistical significance of about
4.9.  Taking into account the systematic uncertainty, we still obtain
a significance of 4.

\begin{table}
\begin{center}
\begin{tabular}{c|c|c|c|c||c|c|c}
& $WZ$ & $ZZ$ & $t\bar t$ & {\bf BP1} & S/B & $\mathcal{S}_{\rm stat}$
  & $\mathcal{S}_{\rm stat\,+\,sys}$\\ \hline \hline 3 leptons &
  111143 & 15282 & 156210 & 4675 & 0.02 & 8.8 & 0.2\\ SFOS & 109093 &
  15102 & 116521 & 4614 & 0.02 & 9.4 & 0.3 \\ $Z$ veto & 15606 & 1969
  & 99826 & 4384 & 0.04 & 12.8 & 0.4\\ \met & 11507 & 871 & 87069 &
  3448 & 0.03 & 10.9 & 0.4 \\ jet veto & 5812 & 298 & 8426 & 1640 &
  0.11 & 13.6 & 1.6
\end{tabular}
\caption{Number of background and signal events for benchmark point
  {\bf BP1} with $M_{\tilde\chi_2^0}=160$ GeV and
  $M_{\tilde\chi_1^0}=100$ GeV after each cut for signal region {\bf
    SRnoZ}. In the last three columns, we present the ratio between
  the number of signal and background events, the statistical
  significance and the significance including systematic errors. All
  numbers are normalized to 300 fb$^{-1}$ at $\sqrt{s}=14$ TeV.}
\label{tab:cutflowSRnoZ}
\end{center}
\end{table}

\begin{table}
\begin{center}
\begin{tabular}{c|c|c|c|c||c|c|c}
& $WZ$ & $ZZ$ & $t\bar t$ & {\bf BP2} & S/B & $\mathcal{S}_{\rm stat}$
  & $\mathcal{S}_{\rm stat\,+\,sys}$\\ \hline \hline 3 leptons &
  111143 & 15282 & 156210 & 132 & 0 & 0.25 & 0.01 \\ SFOS & 109093 &
  15102 & 116521 & 129 & 0 & 0.26 & 0.01 \\ $Z$ request & 93487 &
  13133 & 16694 & 106 & 0 & 0.3 & 0.01\\ $b$-jet veto & 89233 & 12184
  & 6457 & 97 & 0 & 0.3 & 0.01 \\ \met & 1872 & 102 & 288 & 68 & 0.03
  & 1.43 & 0.34\\ $m_T$ & 47 & 5 & 52 & 50 & 0.48 & 4.9 & 4.03\\
\end{tabular}
\caption{Number of background and signal events for benchmark point
  {\bf BP2} with $M_{\tilde\chi_2^0}=400$ GeV and
  $M_{\tilde\chi_1^0}=20$ GeV after each cut for signal region {\bf
    SRZ2}. In the last three columns, we present the ratio between the
  number of signal and background events, the statistical significance
  and the significance including systematic errors. All numbers are
  normalized to 300 fb$^{-1}$ at $\sqrt{s}=14$ TeV.}
\label{tab:cutflowSRZ1}
\end{center}
\end{table}

\begin{figure}
\begin{minipage}[t]{0.47\textwidth} \centering
\begin{tikzpicture}
\node (plot) {\includegraphics[height=.8\textwidth,
    width=.8\textwidth,angle=0] {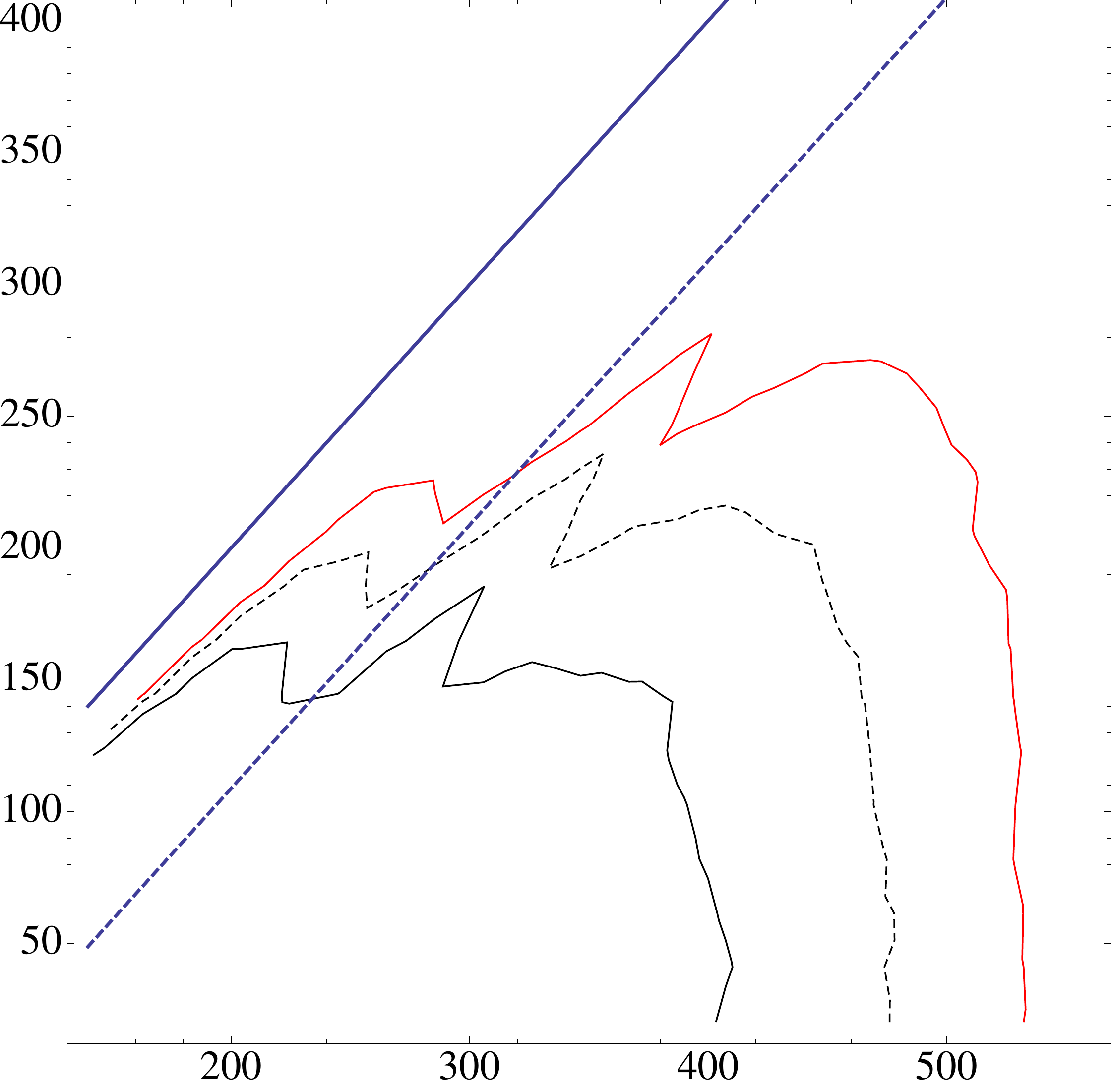}}; \node
      [rotate=90] at (-4,.5) {\Large{$\mathbf{M_{\tilde{\chi}^0_1}~
            [GeV]}$}}; \node at (0,-4) {\Large{$\mathbf{
            M_{\tilde{\chi}^0_2}~ [GeV]}$}};
\end{tikzpicture}
\caption{\em \small Signal significance (only statistical errors) in the
  $\tilde\chi_2^0$-$\tilde\chi_1^0$ mass plane assuming an
  integrated luminosity of 300 fb$^{-1}$ at $\sqrt{s}=14$ TeV. The red, black dashed and black solid
  curve corresponds to 2, 3 and 5$\sigma$, respectively. The blue dashed line indicates
  $M_{\tilde\chi_2^0}-M_{\tilde\chi_1^0}=m_Z$ whereas the blue solid
  curve delimits the region with a singlino LSP.} 
                 \label{exclnosys}
\end{minipage}
\hspace{5mm}
\begin{minipage}[t]{0.47\textwidth} \centering
\begin{tikzpicture}
\node (plot) {\includegraphics[height=.8\textwidth, width=.8\textwidth,angle=0]
             {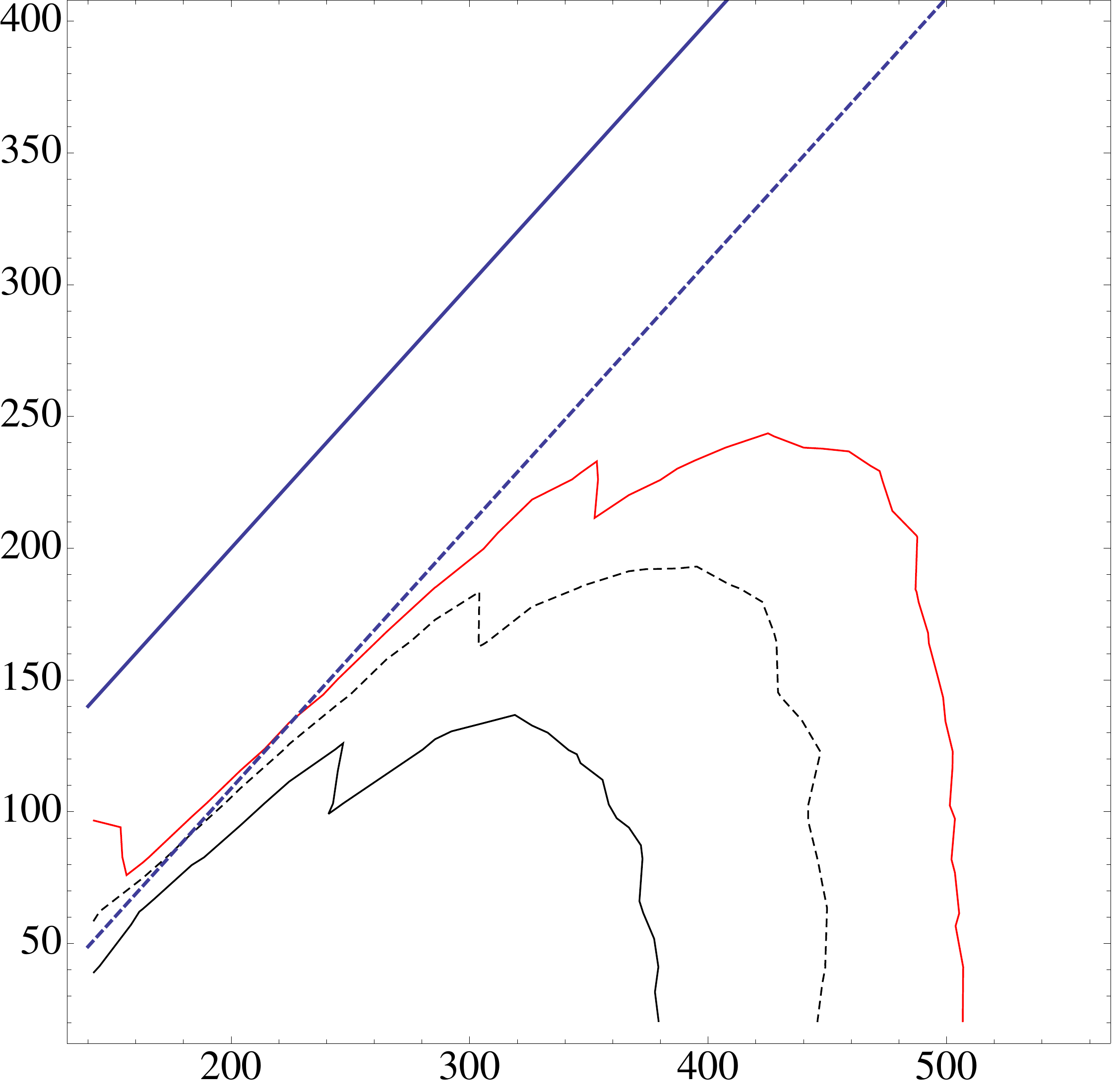}};
\node [rotate=90] at (-4,.5) {\Large{$\mathbf{M_{\tilde{\chi}^0_1}~ [GeV]}$}};
\node at (0,-4) {\Large{$\mathbf{M_{\tilde{\chi}^0_2}~ [GeV]}$}};
\end{tikzpicture}
	   \caption{\em \small Signal significance  including systematic errors. Everything else is the same as in
	   Figure~\ref{exclnosys}.}  
             \label{exclsys}
\end{minipage}
\end{figure}

In Figure~\ref{exclnosys}, we present the exclusion limits in the
$\tilde\chi_2^0$-$\tilde\chi_1^0$ mass plane at the LHC at
$\sqrt{s}=14$ TeV with 300 fb$^{-1}$. The statistical significance is
estimated with Equation~(\ref{eq:sig_stat}).  The best signal region
is chosen for each point in the mass plane. The red, black dashed and
black solid curve correspond to 2$\sigma$, 3$\sigma$ and 5$\sigma$,
respectively. Above the blue solid line, the decay of the higgsino
into a singlino $+\,X$ is not allowed. The blue dashed line
corresponds to $M_{\tilde\chi_2^0}-M_{\tilde\chi_1^0}=m_Z$. Below the
blue dashed line, the $\tilde\chi_2^0$ decays in a 2-body decay with
$\tilde \chi_2^0\rightarrow\tilde\chi_1^0\,Z$. Here, the selection
cuts of the $Z$ enriched signal regions provide the best sensitivity
for our signal. Above the blue dashed curve, the $\tilde \chi_2^0$
decays via off-shell $Z^*$ in a three body final state which is
sensitive to the $Z$ depleted signal region. We are sensitive for
higgsino masses up to 540 GeV for massless singlinos. With decreasing
mass difference between the higgsino and the singlino, the
significance drops sharply. Decreasing the mass splitting reduces the
average $p_T$ of the leptons. Thus, the final state leptons becomes
too soft which does not allow to separate our signal from the SM
background. However, these region can be probed in higgsino pair
production in association with a hard jet
\cite{Baer:2014cua,Han:2013usa,Schwaller:2013baa} or a trilepton search with a
relatively hard initial state radiation jet \cite{Gori:2013ala}.

In Figure~\ref{exclsys}, we included the systematical errors for the
calculation of the significance, see Equation~(\ref{eq:sig_sys}).
Higgsino masses up to 500 GeV can be excluded for massless singlinos.
We are not sensitive to small mass differences between the higgsinos
and singlinos due to the large systematic errors of the $WZ$ and
$t\bar t$ backgrounds. However, our estimate of 10$\%$ is quite
conservative and hence Figure~\ref{exclsys} is a rather pessimistic
estimate of the exclusion limit for small mass differences. As more
data is collected, the systematic uncertainties will be much smaller
and one can expect to cover a substantial portion of the DM allowed
region through the trilepton channel. 

\section{Conclusion}\label{sec:conclude}
In this paper, we considered a light {\it higgsino-singlino world
  scenario} with decoupled matter scalars and gauginos in the
NMSSM. There are phenomenological reasons to consider such a split
scenario. The non-observation of supersymmetric particles with a
relatively heavy Higgs provides strict limits on the soft breaking
scale of supersymmetry. However, finetuning arguments favor relatively
light higgsinos. But a light higgsino LSP with multi TeV scalars and
gauginos typically results in a too small relic density in standard
cosmology.  On the other hand, a supersymmetric model with a light
higgsino-singlino sector can provide a viable DM candidate. If the
higgsino is the NLSP with a small splitting to the singlino LSP,
co--annihilation between both sparticles can lead to the correct
relict density. However, for a relative degeneracy between the
higgsino and the singlino, the production of higgsinos is difficult to
detect at the LHC since the decay products of the higgsinos are very
soft. On the other hand, the right amount of the relic density can be
obtained via resonant annihilation with heavy Higgs scalars while
allowing for a large mass splitting between the higgsino and the
singlino.  Another advantage to consider a {\it higgsino-singlino
  world scenario} is that flavor changing neutral current CP violating
processes are suppressed and that the gravitino problem is solved.
Thus motivated, we focused on the production of a higgsino-like
chargino neutralino pair at the LHC. In particular, we considered the
leptonic decay modes which results in the trilepton and missing
transverse energy final state. In this work, we present a collider study of the {\it
  higgsino-singlino world scenario} at the LHC at $\sqrt{s}=14$ TeV
for an integrated luminosity of 300 fb$^{-1}$. We simulated the signal
and the most important SM backgrounds with recent MC simulations and
we also estimated the detector response with a fast detecter
simulation. We considered three signal regions corresponding to a $Z$
depleted region (for small mass differences between the higgsino and
the singlino) and two $Z$ enriched signal region. We discussed in
detail the cuts for two benchmark scenarios.  We examined the
discovery reach in the higgsino-singlino mass plane. For massless
singlinos, higgsinos with masses up to 500 GeV can be excluded for an
integrated luminosity of 300 fb$^{-1}$ at $\sqrt{s}=14$ TeV. However,
the discovery reach is severely constrained in the small splitting
region due to the low efficiency of the selection cuts and the
assumptions on the systematic errors. Higgsino masses with a mass
splitting of the order of the $Z$ mass boson can be excluded with 200
(300) GeV assuming a systematic error of 0$\%$ (10$\%$). However, the region of small splitting
would require more involved search strategies~\cite{Baer:2014cua,Han:2013usa,Schwaller:2013baa,Gori:2013ala} 
to be accessible at the LHC.
Finally, our results of the discovery reach are also true if we allow for 
non-split scenarios, e.g. if matter scalars are kinematically accessible at the LHC, but do not alter our assumptions of the
decay chain.

\noindent {\bf Acknowledgments:}~ We thank Biplob Bhattacherjee, James
Barnard and Daniel Schmeier for discussions. The research of TSR is
supported by the Australian Research Council.  The work of JSK has
been partially supported by the MICINN, Spain, under contract FPA2010-
17747; Consolider-Ingenio CPAN CSD2007-00042. JSK also thanks the
Spanish MINECO Centro de excelencia Severo Ochoa Program under grant
SEV-2012-0249.


\end{document}